# Electronic Correlation and Pseudogap-like Behavior of High-Temperature Superconductor $La_3Ni_2O_7$

Yidian Li[1*], Xian Du[1*], Yantao Cao[2,5*], Cuiying Pei[3*], Mingxin Zhang[3], Wenxuan Zhao[1], Kaiyi Zhai[1], Runzhe Xu[1], Zhongkai Liu[3,4], Zhiwei Li[2], Jinkui Zhao[5], Gang Li[3], Yanpeng Qi[3,4†], Hanjie Guo[5†], Yulin Chen[3,4,6†], and Lexian Yang[1,7,8†]

[1]*State Key Laboratory of Low Dimensional Quantum Physics, Department of Physics, Tsinghua University, Beijing 100084, China.*
[2]*Key Lab for Magnetism and Magnetic Materials of the Ministry of Education, Lanzhou University, Lanzhou 730000, China*
[3]*School of Physical Science and Technology, ShanghaiTech University and CAS-Shanghai Science Research Center, Shanghai 201210, China.*
[4]*ShanghaiTech Laboratory for Topological Physics, Shanghai 200031, China.*
[5]*Songshan Lake Materials Laboratory, Dongguan, Guangdong 523808, China*
[6]*Department of Physics, Clarendon Laboratory, University of Oxford, Parks Road, Oxford OX1 3PU, UK.*
[7]*Frontier Science Center for Quantum Information, Beijing 100084, China.*
[8]*Collaborative Innovation Center of Quantum Matter, Beijing 100084, China*
*These authors contributed equally to this work.
Emails: YPQ: qiyp@shanghaitech.edu.cn; HJG: hjguo@sslab.org.cn; YLC: yulin.chen@physics.ox.ac.uk; LXY: lxyang@tsinghua.edu.cn

High-temperature superconductivity (HTSC) remains one of the most challenging and fascinating mysteries in condensed matter physics. Recently, superconductivity with transition temperature exceeding liquid-nitrogen temperature is discovered in $La_3Ni_2O_7$ at high pressure, which provides a new platform to explore the unconventional HTSC. In this work, using high-resolution angle-resolved photoemission spectroscopy and *ab-initio* calculation, we systematically investigate the electronic structures of $La_3Ni_2O_7$ at ambient pressure. Our experiments are in nice agreement with *ab-initio* calculations after considering an orbital-dependent band renormalization effect. The strong electron correlation effect pushes a flat band of $d_{z^2}$ orbital component below the Fermi level ($E_F$), which is predicted to locate right at $E_F$ under high pressure. Moreover, the $d_{x^2-y^2}$ band shows a pseudogap-like behavior with suppressed spectral weight and diminished quasiparticle peak near $E_F$. Our findings provide important insights into the electronic structure of $La_3Ni_2O_7$, which will shed light on the understanding of the unconventional superconductivity in nickelates.

Being a long-standing problem yet to be resolved, high-temperature superconductivity (HTSC) exhibits important scientific significance and application prospects [1-3]. Up to date, there are only two families of compounds featuring unconventional superconductivity well above the McMillan limit at ambient pressure: cuprates and iron-based superconductors (Fe-SCs) [1-3]. Both of them show layered crystal structure; the electronic states near the Fermi level ($E_F$) are dominated by $3d$ electrons with strong electronic correlation [1-4]; and the superconductivity is in proximity to a magnetic phase in their phase diagrams [5,6]. However, prominent differences also exist between these two families of superconductors. Cuprates based on $3d^9$ electronic configuration of $Cu^{2+}$ have a single band crossing $E_F$, which is usually described by the one-band Hubbard model [7,8]. By contrast, iron-based superconductors with Fe $3d^6$ electronic configuration have a multi-band and multi-orbital nature [5]. Moreover, cuprate shows a pseudogap above the transition temperature [6], while Fe-SCs are usually bad metals in the normal state. To understand the importance of these similarities and differences in unconventional superconductivity, it is highly desired to explore new materials with HTSC.

Infinite-layer nickel oxides are considered to exhibit unconventional superconductivity despite their relatively low critical temperature [9-13]. Similar to the cuprates and Fe-SCs, $3d$ electrons of Ni dominate the electronic states near $E_F$. Electronic correlation also plays an important role in the electronic structure and superconductivity of Ni-based superconductors (Ni-SCs) [14]. In particular, $Ni^+$ exhibits an effective electronic configuration of $3d^9$, the same as $Cu^{2+}$ in cuprates [15,16], making infinite-layer nickelates an attractive platform for exploring HTSC. Superconductivity at 15 K and 30 K has been observed in thin films of infinite-layer nickelates at ambient and high pressure respectively [17,18]. Superconductivity at 13 K is also observed in the quintuple-layer $Nd_6Ni_5O_{12}$, in which Ni exhibits an effective electronic configuration of $d^{8.8}$ [19]. However, the synthesis of these nickelates is challenging and the superconductivity is realized only in thin films below the McMillan limit regardless of the application of pressure.

Recently, HTSC with $T_c$ up to 80 K was discovered in La$_3$Ni$_2$O$_7$ bulk compound under pressure between 14.0 GPa and 43.5 GPa [20-23]. It belongs to the series of La$_{n+1}$Ni$_n$O$_{3n+1}$ Ruddlesden-Popper phase that can be viewed as the parent phase of the infinite-layer and quintuple-layer nickelates. Many theoretical understandings of the electronic structure and superconductivity of the system have been proposed [24-35]. It was proposed that the Ni ions with an average valence of +2.5 are pressurized to resemble the electronic configuration of Cu$^{2+}$ in cuprates, which favors the superconductivity at high pressure [20]. Simultaneously, the crystal and electronic structures are strongly modulated by the pressure-induced Jahn-Teller distortion. The $3d_{x^2-y^2}$ and $3d_{z^2}$ orbitals strongly interact with O $p$ orbitals, forming intra- and inter-layer σ bonds respectively. The interaction between the occupied Ni $d_{x^2-y^2}$ and O $p$ orbitals also induces the Zhang-Rice singlet in the superconducting state [20,33]. While it is extremely challenging to experimentally investigate the electronic structure at high pressure, the experiments at ambient pressure will nevertheless provide crucial information for its normal-state properties, including the Fermiology, the electronic correlation effect, density wave orders, and non-Fermi-liquid behaviors [36-39]. A previous measurement using angle-resolved photoemission spectroscopy (ARPES) revealed orbital dependent correlation effect in La$_3$Ni$_2$O$_7$ at ambient pressure [40] with more details awaiting further experimental exploration.

In this work, by combining ARPES and *ab-initio* calculation, we systematically investigate the electronic structure of La$_3$Ni$_2$O$_7$ at ambient pressure. The band structure shows a multi-band and multi-orbital nature with Ni $3d_{x^2-y^2}$ and $3d_{z^2}$ orbitals dominating the electronic states near $E_F$ [24]. After being renormalized by 2.7 and 5.0 for the $d_{x^2-y^2}$ and $d_{z^2}$ bands respectively [40], the calculated results are in nice agreement with the experiments, suggesting a strong electronic correlation in the system [26-28,34,39]. In particular, the $3d_{z^2}$ orbitals form a flat band just below $E_F$, which contributes significantly to the electronic density of states (DOS). With high pressure, our *ab-initio* calculation suggests that the band width of the La$_3$Ni$_2$O$_7$ increases and the flat band locates right at $E_F$, which may favor the pressurized HTSC. Furthermore, we observe diminished quasiparticle peak and suppressed spectral weight as the $d_{x^2-y^2}$ band dispersing approaching $E_F$, which shows a pseudogap-like behavior with decreasing temperature and is reminiscent of the normal-state behaviors of cuprates and Fe-SCs [8,41]. Our findings will help construct a unified picture of unconventional HTSC in different families of materials.

As schematically shown in Fig. 1a, $La_3Ni_2O_7$ crystallizes into a layered orthorhombic structure in the space group *Amam* at ambient pressure with lattice constants $a$ = 5.393 Å, $b$ = 5.449 Å, and $c$ = 20.518 Å [20,42]. In the conventional unit cell, corner-sharing $NiO_6$ octahedrons form Ni-O bilayers stacking along the $c$ direction, which are separated by La-O layers. At ambient pressure, the $NiO_6$ octahedrons slightly distort with their vertical axis canting for about 6° away from the $c$ axis (Fig. 1b) [20]. The application of high pressure aligns the octahedrons along $c$ and drives the system into the space group *Fmmm* (Fig. 1c). The Ni-O planes also become flat at high pressure. Within each Ni-O bilayer, the corner-sharing of $NiO_6$ octahedrons through the apical oxygen atoms provides the structural basis for the interlayer σ bonds between Ni $3d_{z^2}$ and O $p$ orbitals[20]. The crystals can be easily cleaved between the bilayers to expose the La-O termination as indicated by the blue planes in Fig. 1a. The Brillouin zones (BZs) corresponding to the primitive and conventional unit cells are shown in Fig. 1d with high-symmetry points of the conventional BZ indicated.

Our sample shows clear Laue peaks in Fig. 1e, confirming its orthorhombic structure. The metallic temperature dependence of the resistance suggests the proper element composition, particularly of the oxygen, since a slight deficiency of oxygen will drive the system into an insulating phase [43]. We note slight kink-like features in the resistance curve, which were previously associated to a possible charge-density-wave (CDW) transition [22,44]. At the pressure of 27.9 GPa, our samples show a superconducting transition with onset critical temperature of about 80 K, in good agreement with the previous measurements [20,45].

Figures 2a and b show the *ab-initio* calculation of the band structure at the ambient pressure with onsite Coulomb interaction on Ni atoms $U = 0$ eV and 4 eV respectively. The onsite Coulomb interaction expels the $t_{2g}$ bands towards higher binding energies, so that the $e_g$ and $t_{2g}$ bands do not hybridize for $U = 4$ eV. Simultaneously, the δ band with a flat band top submerges below $E_F$. Fig. 2c shows the calculated three-dimensional (3D) Fermi surface (FS), which suggests a weak $k_z$ dispersion, consistent with our experiments at different photon energies (See the Supplementary). There are three sets of FS sheets in the conventional BZ: a cylindrical electron pocket around the $\bar{\Gamma}$ point, a cylindrical hole pocket around the $\bar{X}/\bar{Y}$ point, and a flat FS sheet near the $\bar{S}$ point (Figs. 2c and d). In the experiment, we observe a large rectangular pocket around the $\bar{\Gamma}$ point (Fig. 2e), which originates from the band folding induced by the distortion (Figs. 1a-c) of the NiO$_6$ octahedrons at the ambient pressure (See the Supplementary). In the conventional BZ (orange lines) (Fig. 2f), we resolve a hole pocket around the $\bar{X}/\bar{Y}$ point and a small square-like pocket around the $\bar{S}$ point, agreeing nicely with the calculation. The parallel FS provides an ideal condition for FS nesting (See the Supplementary), which may drive an instability towards long-range ordering states [22], as suggested by the calculation of the spin susceptibility [46,47]. However, the existence of the density-wave ordering and its origin require further experimental and theoretical investigations. The electron pocket around $\bar{\Gamma}$ is better visualized using laser-ARPES, consistent with the previous report [40]. Based on the experimental FS in Figs. 2e-g, the carrier density is estimated to be about 0.249 e$^-$ per unit cell, in nice agreement with the theoretical predictions [20]. Fig. 2h shows the band structure measured in a large energy range. The most prominent spectral features locate between -5.0 and -1.5 eV, with relatively weak intensity near $E_F$. The comparison between the integrated energy-distribution curve (EDC) and calculated DOS shows an overall consistence (Fig. 2i).

Figure 3 investigates the fine electronic structure near $E_F$. The 3D volume plot in Fig. 3a shows the overall band dispersion along different momentum directions. We observe a hole band δ with a flat band top and an electron band α/β around the $\bar{\Gamma}'$ point (the center of the second Brillouin zone) (Figs. 3b and c). The flat band is very close to $E_F$, compared to that at about 50 meV below $E_F$ in the previous ARPES measurements [48]. This difference might be due to the oxygen vacancies in the samples. The electron band α/β was also observed in the second BZ and at other photon energies (See the Supplementary). Near the $\bar{S}$ point, we observe an electron band γ dispersing to about 100 meV below $E_F$ (Figs. 3b and d). Data collected using 75 eV and 86 eV are in general similar to each other, consistent with the layered crystal structure (Figs. 3c and e). It is noteworthy that the experimental band structure of $La_3Ni_2O_7$ is very similar to that of $La_4Ni_3O_{10}$ [49], which also shows superconductivity at high pressure [50-52].

The calculated band structure with $U = 4$ eV is renormalized by a factor of about 5 and overlaid on the ARPES spectra in Figs. 3b-e. The calculation nicely reproduces the experimental band structure, suggesting a strong electronic correlation in the system [28,39]. Consistently, we notice a waterfall-like structure at high energies, which has been widely observed in various correlated superconductors (white arrow in Fig. 3b) [53,54]. It is noteworthy that the α/β and γ bands, with the estimated renormalization factor of about 2.7, are less-correlated than the flat δ band, in agreement with the orbital-dependent electronic correlation effect in the previous report [46,48], as indicated by the white dashed lines in Figs. 3b, 3d and 3e. By comparing with the orbital-projected calculation in Fig. 3f (See the Supplementary), we identify that the highly dispersive α/β band and the flat δ band are contributed by Ni $3d_{x^2-y^2}$ and $3d_{z^2}$ orbitals, respectively [25,27,40,48]. The electronic correlation and its orbital-dependence agree with previous ARPES experiment, while the latter reported a further enhancement of the correlation effect of the $3d_{z^2}$ band along the $\bar{\Gamma}\bar{S}$ direction [40].

The nice agreement between the experiments and calculations confirms the validity of our *ab-initio* calculations in the investigation of the electronic structure of $La_3Ni_2O_7$. We further calculated the band structure at high pressure that flattens the Ni-O layer and aligns $NiO_6$ octahedrons along *c*. As shown in Fig. 3g, the bandwidth increases by about 25% and the δ band (the flat band in the experiment) moves closer to $E_F$. Moreover, a van Hove singularity (saddle point) at the $\bar{X}$ point also moves closer to $E_F$. Along $\bar{\Gamma}\bar{S}$ the $d_{x^2-y^2}$ and $d_{z^2}$ orbitals show an enhanced mixture at high pressure. These changes of the band structure at high pressure will enhance the electronic DOS near $E_F$, which favors the superconducting transition.

Figure 4 investigates the temperature evolution of the electronic structure. With increasing temperature, the flat band stays put at temperatures up to about 120 K (Fig. 4a), as can be seen from the EDCs at $\bar{\Gamma}'$ (Fig. 4b). We notice that the spectral weight of the α/β band gradually reduces as approaching $E_F$ without a clear quasiparticle peak near $E_F$, which is also noticeable in Ref. [40] and is beyond the expectation of the typical Fermi-liquid picture. This behavior is better visualized in the laser-based ARPES measurements with improved energy and momentum resolutions as shown in Fig. 4c. In a large temperature range from 19 K to 120 K, the spectral weight of the α band remains strongly suppressed near $E_F$. The degree of the suppression reduces with increasing temperature, which is evident by the gradual shift of the leading edge of the EDCs at $k_F$ (Fig. 4b) towards high energies. The leading edge is below $E_F$ even at 120 K, which may be related to the multiple kink-like features in the resistance that are related to the density-wave states [36-39,55] (Fig. 1f). Fig. 4d plots the leading-edge position as a function of temperature. It is clear that the leading-edge shift towards high-energies starts at about 100 K and reaches its maximum of about 9 meV at 19 K, showing a Bardeen-Cooper-Shrieffer (BCS)-like evolution.

The temperature dependence of the $d_{x^2-y^2}$ band is reminiscent of the pseudogap in cuprate superconductors [6,8], However, one should notice that the observed pseudogap-like behavior is not necessarily related to the HTSC under pressure. It is also noteworthy that recent theoretical calculations likewise unravel a pseudogap in the DOS near $E_F$ [56], in agreement with our experiments. We suggest that the observed pseudogap-like behavior of $La_3Ni_2O_7$ may be due to the strong electronic correlation, the formation of possible density wave, and/or the hybridization between the "localized" $d_{z^2}$ orbital and "itinerant" $d_{x^2-y^2}$ orbitals, which proposes further experimental and theoretical studies. On the other hand, the lacking of a well-defined Fermi edge in a metallic system suggests a non-Fermi-liquid behavior. Consistent with our observation, a non-Fermi-liquid behavior has also been revealed by the optical study [39] and our ultrafast reflectivity measurements [57]. Moreover, a non-Fermi-liquid state has been theoretically predicted in the *Fmmm* phase of $La_3Ni_2O_7$ at high pressure, which is argued to be intimately correlated to the superconducting state [26].

In conclusion, we have comprehensively studied the electronic structure of $La_3Ni_2O_7$ at ambient pressure. The nice agreement between experiment and calculation with onsite Coulomb interaction and band renormalization suggests an important role of electronic correlation in the system. The $d_{z^2}$ band is flattened by the strong correlation, while the less-correlated $d_{x^2-y^2}$ band shows a pseudogap-like behavior, which are reminiscent of the orbital-selective band renormalization in Fe-SCs and pseudogap in cuprates. At high pressure, the flat band is predicted to locate right at $E_F$ and is believed to be crucial for the HTSC. Our systematic experiments provide an electronic basis for exploring the Ni-based unconventional HTSC.

**METHODS**

**Sample growth**

Single crystals of $La_3Ni_2O_7$ were grown by the high-pressure optical floating zone technique [44]. Raw materials of $La_2O_3$ were dried at 900 ºC overnight prior to the reaction to avoid moisture contamination. Subsequently, $La_2O_3$ and NiO powders were mixed in a stoichiometric ratio with 1% ~2% excess of NiO and then ground thoroughly. The mixture was then pressed into pellets and sintered at 1100 °C for 2 days with several intermediate grindings. The reactant was pressed into seed and feed rods with a diameter of about 6 mm under hydrostatic pressure and then sintered at 1300 °C for 2 hours. The single-crystal growth was performed in a high-pressure optical floating zone furnace (HKZ-300) with an oxygen pressure of 15 bar. The growth rate was 2 mm/h and the feed and seed rods were rotated in opposite directions with speeds of 20 and 25 rpm, respectively.

**High-pressure transport measurement**

High-pressure in-situ electrical transport property was performed in Physical Property Measurement System (PPMS-9T) as described elsewhere [58-60]. Nonmagnetic BeCu diamond anvil cell (DAC) with 200 um culet was employed to perform the in situ high-pressure resistivity measurements. A cubic-BN/epoxy mixture covers BeCu gasket as an insulator layer. The van der Pauw four-probe electrodes were applied on the cleavage plane. The pressure was determined by ruby fluorescence method [61].

**ARPES**

Synchrotron-based ARPES measurements were conducted at beamline 03U in Shanghai Synchrotron Radiation Facility (SSRF). The samples were cleaved *in-situ* under ultra-high vacuum below $7\times 10^{-11}$ mbar. Data were collected with a Scienta DA30 electron analyzer. The total energy and angular resolutions were set to 20 meV and 0.2°, respectively.

Laser-based ARPES measurements were conducted at Tsinghua University. The 7-eV laser was generated by frequency doubling in a KBBF crystal and focused on the sample by an optical lens. Samples were cleaved *in-situ* under ultra-high vacuum below $6\times 10^{-11}$ mbar. Data were collected with a Scienta DA30L electron analyzer. The total energy and angular resolutions were set to 3 meV and 0.2°, respectively.

**Ab-initio calculations**

First-principles band structure calculations were performed using Vienna *ab-initio* simulation package (VASP) [62] with a plane wave basis. The exchange-correlation energy was considered under Perdew-Burke-Ernzerhof (PBE) type generalized gradient approximation (GGA) [63] with spin-orbit coupling included. Hubbard $U = 4$ eV was applied to describe the localized $3d$ orbitals of Ni atoms. The cutoff energy for the plane-wave basis was set to 400 eV. A Γ-centered Monkhorst-Pack $k$-point mesh of 19×19×5 was adopted in the self-consistent calculations. Based on the tight-binding type Hamiltonian constructed from maximally localized Wannier functions (MLWF) supplied by the Wannier90 code [64], the surface-projected band structures were calculated with the WANNIERTOOLS package [65].

**ACKNOWLEDGEMENTS**

This work is funded by the National Key R&D Program of China (Grants No. 2022YFA1403100 and No. 2022YFA1403200), the National Natural Science Foundation of China (Grants No. 12275148, No. 12004270, and No. 52272265), and Guangdong Basic and Applied Basic Research Foundation (Grant No. 2022B1515120020). L. X. Y. acknowledges the support from Tsinghua University Initiative Scientific Research Program.

**AUTHOR CONTRIBUTIONS**

L.X.Y. conceived the scientific project. Y.D.L. and X.D. carried out ARPES measurements and data analyses with the assistance of W.X.Z., K.Y.Z., R.Z.X., Z.K.L., and Y.L.C. X.D. performed the *ab-initio* calculations with the help from G.L. Single crystals were synthesized and characterized by Y.T.C., M.X.Z., C.Y.P., Z.W.L., J.K.Z., Y.P.Q., and H.J.G. All authors contributed to the scientific planning and discussion.

# REFERENCES

[1]  B. Keimer, S. A. Kivelson, M. R. Norman, S. Uchida, and J. Zaanen, *From quantum matter to high-temperature superconductivity in copper oxides*, Nature **518**, 179 (2015).

[2]  X. Zhou, W.-S. Lee, M. Imada, N. Trivedi, P. Phillips, H.-Y. Kee, P. Törmä, and M. Eremets, *High-temperature superconductivity*, Nat. Rev. Phys. **3**, 462 (2021).

[3]  Q. Si, R. Yu, and E. Abrahams, *High-temperature superconductivity in iron pnictides and chalcogenides*, Nat. Rev. Mater. **1**, 16017 (2016).

[4]  C. C. Tsuei and J. R. Kirtley, *Pairing symmetry in cuprate superconductors*, Rev. Mod. Phys. **72**, 969 (2000).

[5]  J. Paglione and R. L. Greene, *High-temperature superconductivity in iron-based materials*, Nat. Phys. **6**, 645 (2010).

[6]  M. Hashimoto, I. M. Vishik, R.-H. He, T. P. Devereaux, and Z.-X. Shen, *Energy gaps in high-transition-temperature cuprate superconductors*, Nat. Phys. **10**, 483 (2014).

[7]  E. Dagotto, *Correlated electrons in high-temperature superconductors*, Rev. Mod. Phys. **66**, 763 (1994).

[8]  A. Damascelli, Z. Hussain, and Z.-X. Shen, *Angle-resolved photoemission studies of the cuprate superconductors*, Rev. Mod. Phys. **75**, 473 (2003).

[9]  X. Zhou, P. Qin, Z. Feng, H. Yan, X. Wang, H. Chen, Z. Meng, and Z. Liu, *Experimental progress on the emergent infinite-layer Ni-based superconductors*, Mater. Today **55**, 170 (2022).

[10] D. Li, K. Lee, B. Y. Wang, M. Osada, S. Crossley, H. R. Lee, Y. Cui, Y. Hikita, and H. Y. Hwang, *Superconductivity in an infinite-layer nickelate*, Nature **572**, 624 (2019).

[11] W. E. Pickett, *The dawn of the nickel age of superconductivity*, Nat. Rev. Phys. **3**, 7 (2021).

[12] D. Li, B. Y. Wang, K. Lee, S. P. Harvey, M. Osada, B. H. Goodge, L. F. Kourkoutis, and H. Y. Hwang, *Superconducting Dome in $Nd_{1-x}Sr_xNiO_2$ Infinite Layer Films*, Phys. Rev. Lett. **125**, 027001 (2020).

[13] M. Osada, B. Y. Wang, B. H. Goodge, S. P. Harvey, K. Lee, D. Li, L. F. Kourkoutis, and H. Y. Hwang, *Nickelate Superconductivity without Rare-Earth Magnetism: (La,Sr)$NiO_2$*, Adv. Mater. **33**, 2104083 (2021).

[14] A. Kreisel, B. M. Andersen, A. T. Rømer, I. M. Eremin, and F. Lechermann, *Superconducting Instabilities in Strongly Correlated Infinite-Layer Nickelates*, Phys. Rev. Lett. **129**, 077002 (2022).

[15] M. Hepting, D. Li, C. J. Jia, H. Lu, E. Paris, Y. Tseng, X. Feng, M. Osada, E. Been, Y. Hikita *et al.*, *Electronic structure of the parent compound of superconducting infinite-layer nickelates*, Nat. Mater. **19**, 381 (2020).

[16] H. Sakakibara, H. Usui, K. Suzuki, T. Kotani, H. Aoki, and K. Kuroki, *Model Construction and a Possibility of Cupratelike Pairing in a New $d^9$ Nickelate Superconductor (Nd,Sr)NiO$_2$*, Phys. Rev. Lett. **125**, 077003 (2020).

[17] S. Zeng, C. Li, L. E. Chow, Y. Cao, Z. Zhang, C. S. Tang, X. Yin, Z. S. Lim, J. Hu, P. Yang *et al.*, *Superconductivity in infinite-layer nickelate La$_{1-x}$Ca$_x$NiO$_2$ thin films*, Sci. Adv. **8**, eabl9927 (2022).

[18] N. N. Wang, M. W. Yang, Z. Yang, K. Y. Chen, H. Zhang, Q. H. Zhang, Z. H. Zhu, Y. Uwatoko, L. Gu, X. L. Dong *et al.*, *Pressure-induced monotonic enhancement of Tc to over 30 K in superconducting Pr$_{0.82}$Sr$_{0.18}$NiO$_2$ thin films*, Nat. Commun. **13**, 4367 (2022).

[19] G. A. Pan, D. Ferenc Segedin, H. LaBollita, Q. Song, E. M. Nica, B. H. Goodge, A. T. Pierce, S. Doyle, S. Novakov, D. Córdova Carrizales *et al.*, *Superconductivity in a quintuple-layer square-planar nickelate*, Nat. Mater. **21**, 160 (2022).

[20] H. Sun, M. Huo, X. Hu, J. Li, Z. Liu, Y. Han, L. Tang, Z. Mao, P. Yang, B. Wang *et al.*, *Signatures of superconductivity near 80 K in a nickelate under high pressure*, Nature **621**, 493 (2023).

[21] J. Hou, P. T. Yang, and J. Y. L. Z. Y. Liu, P. F. Shan, L. Ma, G. Wang, N. N. Wang, H. Z. Guo, J. P. Sun, Y. Uwatoko, M. Wang, G.-M. Zhang, B. S. Wang, J.-G. Cheng, *Emergence of high-temperature superconducting phase in the pressurized La$_3$Ni$_2$O$_7$ crystals*, arXiv: 2307.09865 (2023).

[22] Y. Zhang, D. Su, and H. S. Yanen Huang, Mengwu Huo, Zhaoyang Shan, Kaixin Ye, Zihan Yang, Rui Li, Michael Smidman, Meng Wang, Lin Jiao, Huiqiu Yuan, *High-temperature superconductivity with zero-resistance and strange metal behavior in La$_3$Ni$_2$O$_7$*, arXiv: 2307.14819 (2023).

[23] M. Zhang, C. Pei, and Y. Z. Qi Wang, Changhua Li, Weizheng Cao, Shihao Zhu, Juefei Wu, Yanpeng Qi, *Effects of Pressure and Doping on Ruddlesden-Popper phases La$_{n+1}$Ni$_n$O$_{3n+1}$*, J. Mater. Sci. Technol. **185**, 147 (2024).

[24] Z. Luo, X. Hu, and W. W. Meng Wang, Dao-Xin Yao, *Bilayer two-orbital model of La$_3$Ni$_2$O$_7$ under pressure*, Phys. Rev. Lett. **131**, 126001 (2023).

[25] Y. Zhang, L.-F. Lin, and E. D. Adriana Moreo, *Electronic structure, orbital-selective behavior, and magnetic tendencies in the bilayer nickelate superconductor La$_3$Ni$_2$O$_7$ under pressure*, arXiv: 2306.03231 (2023).

[26] F. Lechermann, J. Gondolf, and I. M. E. Steffen Bötzel, *Electronic correlations and superconducting instability in La₃Ni₂O₇ under high pressure*, *arXiv: 2306.05121* (2023).

[27] Y. Cao and Y.-f. Yang, *Flat bands promoted by Hund's rule coupling in the candidate double-layer high-temperature superconductor La₃Ni₂O₇*, *arXiv: 2307.06806* (2023).

[28] Z. Liao, L. Chen, and Y. W. Guijing Duan, Changle Liu, Rong Yu, Qimiao Si, *Electron correlations and superconductivity in La₃Ni₂O₇ under pressure tuning*, *arXiv: 2307.16697* (2023).

[29] Y. Shen, M. Qin, and G.-M. Zhang, *Effective bi-layer model Hamiltonian and density-matrix renormalization group study for the high-Tc superconductivity in La₃Ni₂O₇ under high pressure*, *arXiv: 2306.07837* (2023).

[30] Y. Gu, C. Le, and X. W. Zhesen Yang, Jiangping Hu, *Effective model and pairing tendency in bilayer Ni-based superconductor La₃Ni₂O₇*, *arXiv: 2306.07275* (2023).

[31] Y.-f. Yang, G.-M. Zhang, and F.-C. Zhang, *Minimal effective model and possible high-Tc mechanism for superconductivity of La₃Ni₂O₇ under high pressure*, *arXiv: 2308.01176* (2023).

[32] Y.-H. Tian, Y. Chen, and R.-Q. H. Jia-Ming Wang, Zhong-Yi Lu, *Correlation Effects and Concomitant Two-Orbital s±-Wave Superconductivity in La₃Ni₂O₇ under High Pressure*, *arXiv: 2308.09698* (2023).

[33] W. Wú, Z. Luo, and M. W. Dao-Xin Yao, *Charge Transfer and Zhang-Rice Singlet Bands in the Nickelate Superconductor La3Ni2O7 under Pressure*, *arXiv: 2307.05662* (2023).

[34] V. Christiansson, F. Petocchi, and P. Werner, *Correlated electronic structure of La₃Ni₂O₇ under pressure*, *arXiv: 2306.07931* (2023).

[35] J.-X. Zhang, H.-K. Zhang, and Z.-Y. W. Yi-Zhuang You, *Strong Pairing Originated from an Emergent Z₂ Berry Phase in La₃Ni₂O₇*, *arXiv:2309.05726* (2023).

[36] J. C. Xiaoyang Chen, Zhicheng Jiang, Jiong Mei, Kun Jiang, Jie Li, Stefano Agrestini, Mirian Garcia-Fernandez, Xing Huang, Hualei Sun, Dawei Shen, Meng Wang, Jiangping Hu, Yi Lu, Ke-Jin Zhou, Donglai Feng, *Electronic and magnetic excitations in La₃Ni₂O₇*, *arXiv:2401.12657* (2023).

[37] Y. Z. Zhao Dan, Mengwu Huo, Yu Wang, Linpeng Nie, Meng Wang, Tao Wu, Xianhui Chen, *Spin-density-wave transition in double-layer nickelate La₃Ni₂O₇*, *arXiv:2402.03952* (2024).

[38] K. Chen, X. Liu, J. Jiao, M. Zou, C. Jiang, X. Li, Y. Luo, Q. Wu, N. Zhang, Y. Guo *et al.*, *Evidence of Spin Density Waves in La₃Ni₂O₇₋δ*, Phys. Rev. Lett. **132**, 256503 (2024).

[39] Z. Liu, M. Huo, and Q. L. Jie Li, Yuecong Liu, Yaomin Dai, Xiaoxiang Zhou, Jiahao Hao, Yi Lu, Meng Wang, Hai-Hu Wen, *Electronic correlations and energy gap in the bilayer nickelate La₃Ni₂O₇*, *arXiv: 2307.02950* (2023).

[40] J. Yang, H. Sun, X. Hu, Y. Xie, T. Miao, H. Luo, H. Chen, B. Liang, W. Zhu, G. Qu *et al.*, *Orbital-dependent electron correlation in double-layer nickelate $La_3Ni_2O_7$*, Nat. Commun. **15**, 4373 (2024).

[41] M. Yi, D. H. Lu, R. Yu, S. C. Riggs, J. H. Chu, B. Lv, Z. K. Liu, M. Lu, Y. T. Cui, M. Hashimoto *et al.*, *Observation of Temperature-Induced Crossover to an Orbital-Selective Mott Phase in $A_xFe_{2-y}Se_2$ (A =K, Rb) Superconductors*, Phys. Rev. Lett. **110**, 067003 (2013).

[42] C. D. Ling, D. N. Argyriou, G. Wu, and J. J. Neumeier, *Neutron Diffraction Study of $La_3Ni_2O_7$: Structural Relationships Among n=1, 2, and 3 Phases $La_{n+1}Ni_nO_{3n+1}$*, J. Solid. State. Chem. **152**, 517 (2000).

[43] S. Taniguchi, T. Nishikawa, Y. Yasui, Y. Kobayashi, J. Takeda, S.-i. Shamoto, and M. Sato, *Transport, Magnetic and Thermal Properties of $La_3Ni_2O_{7-\delta}$*, J. Phys. Soc. Jpn. **64**, 1644 (1995).

[44] Z. Liu, H. Sun, M. Huo, X. Ma, Y. Ji, E. Yi, L. Li, H. Liu, J. Yu, Z. Zhang *et al.*, *Evidence for charge and spin density waves in single crystals of $La_3Ni_2O_7$ and $La_3Ni_2O_6$*, Sci. China Phys. Mech. **66**, 217411 (2022).

[45] The results about the superconductivity at high pressure will be presented somewhere else.

[46] D. A. Shilenko and I. V. Leonov, *Correlated electronic structure, orbital-selective behavior, and magnetic correlations in double-layer $La_3Ni_2O_7$ under pressure*, Phys. Rev. B **108**, 125105 (2023).

[47] K. J. Yuxin Wang, Ziqiang Wang, Fu-Chun Zhang, Jiangping Hu, *Electronic structure and superconductivity in bilayer $La_3Ni_2O_7$*, *arXiv:2401.15097* (2024).

[48] J. Yang, H. Sun, and Y. X. Xunwu Hu, Taimin Miao, Hailan Luo, Hao Chen, Bo Liang, Wenpei Zhu, Gexing Qu, Cui-Qun Chen, Mengwu Huo, Yaobo Huang, Shenjin Zhang, Fengfeng Zhang, Feng Yang, Zhimin Wang, Qinjun Peng, Hanqing Mao, Guodong Liu, Zuyan Xu, Tian Qian, Dao-Xin Yao, Meng Wang, Lin Zhao, X. J. Zhou, *Orbital-Dependent Electron Correlation in Double-Layer Nickelate $La_3Ni_2O_7$*, *arXiv: 2309.01148* (2023).

[49] H. Li, X. Zhou, T. Nummy, J. Zhang, V. Pardo, W. E. Pickett, J. F. Mitchell, and D. S. Dessau, *Fermiology and electron dynamics of trilayer nickelate $La_4Ni_3O_{10}$*, Nat. Commun. **8**, 704 (2017).

[50] M. Zhang, C. Pei, and Y. C. Xian Du, Qi Wang, Juefei Wu, Yidian Li, Yi Zhao, Changhua Li, Weizheng Cao, Shihao Zhu, Qing Zhang, Na Yu, Peihong Cheng, Jinkui Zhao, Yulin Chen, Hanjie Guo, Lexian Yang, Yanpeng Qi, *Superconductivity in trilayer nickelate $La_4Ni_3O_{10}$ under pressure*, *arXiv:2311.07423* (2023).

[51] Y. Zhu, E. Zhang, and X. C. Bingying Pan, Lixing Chen, Huifen Ren, Feiyang Liu, Junjie Wang, Donghan Jia, Hongliang Wo, Yiqing Gu, Yimeng Gu, Li Ji, Wenbin Wang, Huiyang Gou,

Yao Shen, Tianping Ying, Jiangang Guo, Jun Zhao, *Signatures of superconductivity in trilayer La$_4$Ni$_3$O$_{10}$ single crystals*, *arXiv:2311.07353* (2023).

[52] Q. Li, Y.-J. Zhang, and Y. Z. Zhe-Ning Xiang, Xiyu Zhu, Hai-Hu Wen, *Signature of superconductivity in pressurized La$_4$Ni$_3$O$_{10}$*, *arXiv:2311.05453* (2023).

[53] J. S. Zhou, R. Z. Xu, X. Q. Yu, F. J. Cheng, W. X. Zhao, X. Du, S. Z. Wang, Q. Q. Zhang, X. Gu, S. M. He *et al.*, *Evidence for Band Renormalizations in Strong-Coupling Superconducting Alkali-Fulleride Films*, Phys. Rev. Lett. **130**, 216004 (2023).

[54] B. P. Xie, K. Yang, D. W. Shen, J. F. Zhao, H. W. Ou, J. Wei, S. Y. Gu, M. Arita, S. Qiao, H. Namatame *et al.*, *High-Energy Scale Revival and Giant Kink in the Dispersion of a Cuprate Superconductor*, Phys. Rev. Lett. **98**, 147001 (2007).

[55] T. J. H. Rustem Khasanov, Dariusz J. Gawryluk, Loïc Pierre Sorel, Steffen Bötzel, Frank Lechermann, Ilya M. Eremin, Hubertus Luetkens, Zurab Guguchia, *Pressure-Induced Split of the Density Wave Transitions in La$_3$Ni$_2$O$_{7-\delta}$*, *arXIv:2402.10485* (2024).

[56] S. B. o. Frank Lechermann, and Ilya M. Eremin, *Electronic instability, layer selectivity and Fermi arcs in La$_3$Ni$_2$O$_7$*, arXiv:2403.12831 (2024).

[57] Y. Li, Y. Cao, and P. P. L. Y. Liu, H. Lin, Cuiying Pei, Mingxin Zhang, H. Wu, Xian Du, Wenxuan Zhao, Kaiyi Zhai, J. K. Zhao, M. -L. Lin, Pingheng Tan, Yanpeng Qi, Gang Li, Hanjie Guo, Luyi Yang, and Lexian Yang, *Ultrafast Dynamics of Bilayer and Trilayer Nickelate Superconductors*, *arXiv:2403.05012* (2024).

[58] C. Pei, J. Zhang, Q. Wang, Y. Zhao, L. Gao, C. Gong, S. Tian, R. Luo, M. Li, W. Yang *et al.*, *Pressure induced superconductivity at 32 K in MoB$_2$*, Natl. Sci. Rev. **10**, nwad034 (2023).

[59] C. Pei, T. Ying, Q. Zhang, X. Wu, T. Yu, Y. Zhao, L. Gao, C. Li, W. Cao, Q. Zhang *et al.*, *Caging-Pnictogen-Induced superconductivity in skutterudites IrX$_3$ (X = As, P)*, J. Am. Chem. Soc. **144**, 6208 (2022).

[60] Q. Wang, P. Kong, W. Shi, C. Pei, C. Wen, L. Gao, Y. Zhao, Q. Yin, Y. Wu, G. Li *et al.*, *Charge density wave orders and enhanced superconductivity under pressure in the kagome metal CsV$_3$Sb$_5$*, Adv. Mater. **33**, e2102813 (2021).

[61] H. K. Mao, J. Xu, and P. M. Bell, *Calibration of the ruby pressure gauge to 800 kbar under quasi-hydrostatic conditions*, J. Geophys. Res. **91**, 4673 (1986).

[62] G. Kresse and J. Furthmüller, *Efficient iterative schemes for ab initio total-energy calculations using a plane-wave basis set*, Phys. Rev. B **54**, 11169 (1996).

[63] J. P. Perdew, K. Burke, and M. Ernzerhof, *Generalized Gradient Approximation Made Simple*, Phys. Rev. Lett. **77**, 3865 (1996).

[64] A. A. Mostofi, J. R. Yates, and I. S. Y.-S. Lee, D. Vanderbilt, N. Marzari, *wannier90: A tool for obtaining maximally-localised Wannier functions.*, Comput. Phys. Commun. **178**, 685 (2008).

[65] Q. Wu, S. Zhang, H.-F. Song, M. Troyer, and A. A. Soluyanov, *WannierTools: An open-source software package for novel topological materials*, Comput. Phys. Commun. **224**, 405 (2018).

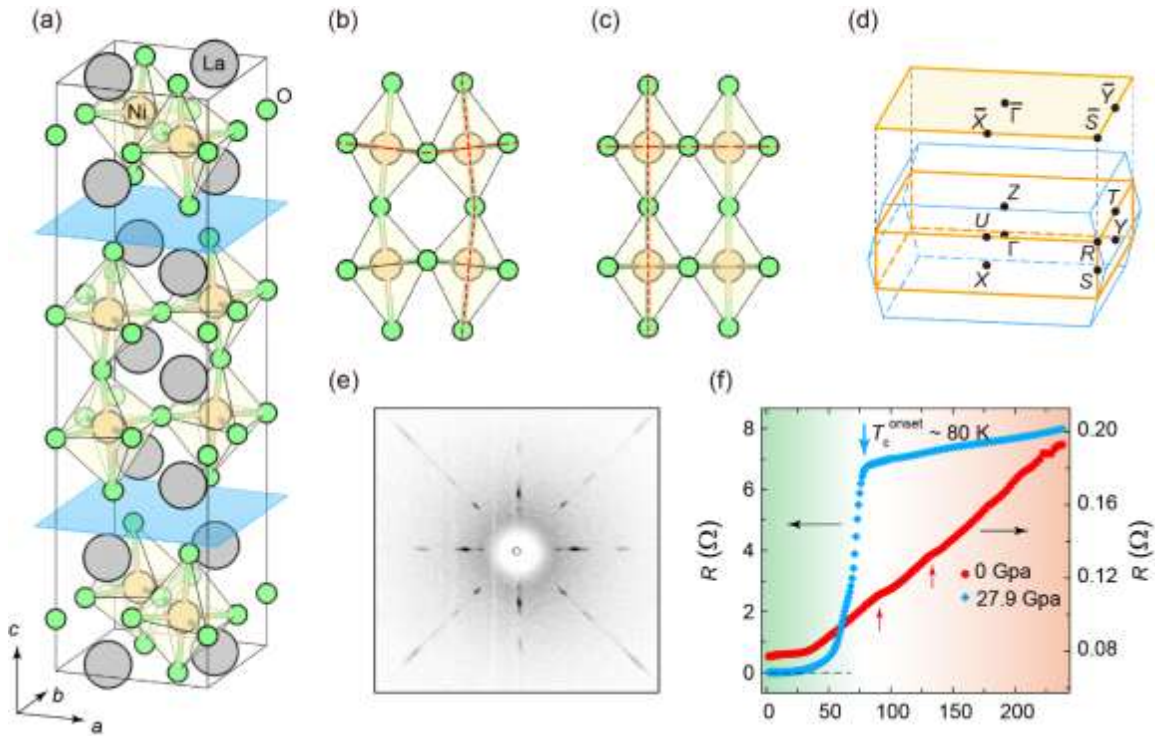

**Fig. 1.** Basic properties of $La_3Ni_2O_7$. (a) Schematic illustration of the crystal structure of $La_3Ni_2O_7$. The blue plane indicates the cleavage plane. (b, c) The side view of the $NiO_6$ bilayer at ambient (b) and high (c) pressure. The red lines indicate the distortion and alignment of the octahedrons at ambient and high pressure. (d) Brillouin zone (BZ) of the conventional (orange lines) and primitive (blue lines) unit cells at ambient pressure. Surface projection and high symmetry points of the conventional BZ are also shown. (e) Laue pattern showing the orthorhombic structure of the crystal. (f) Resistance as a function of temperature at ambient and high-pressure. Superconducting transition under the pressure of 27.9 GPa at about 80 K is indicated by the blue arrow. The red arrows indicate the kink-like features in the resistance.
18

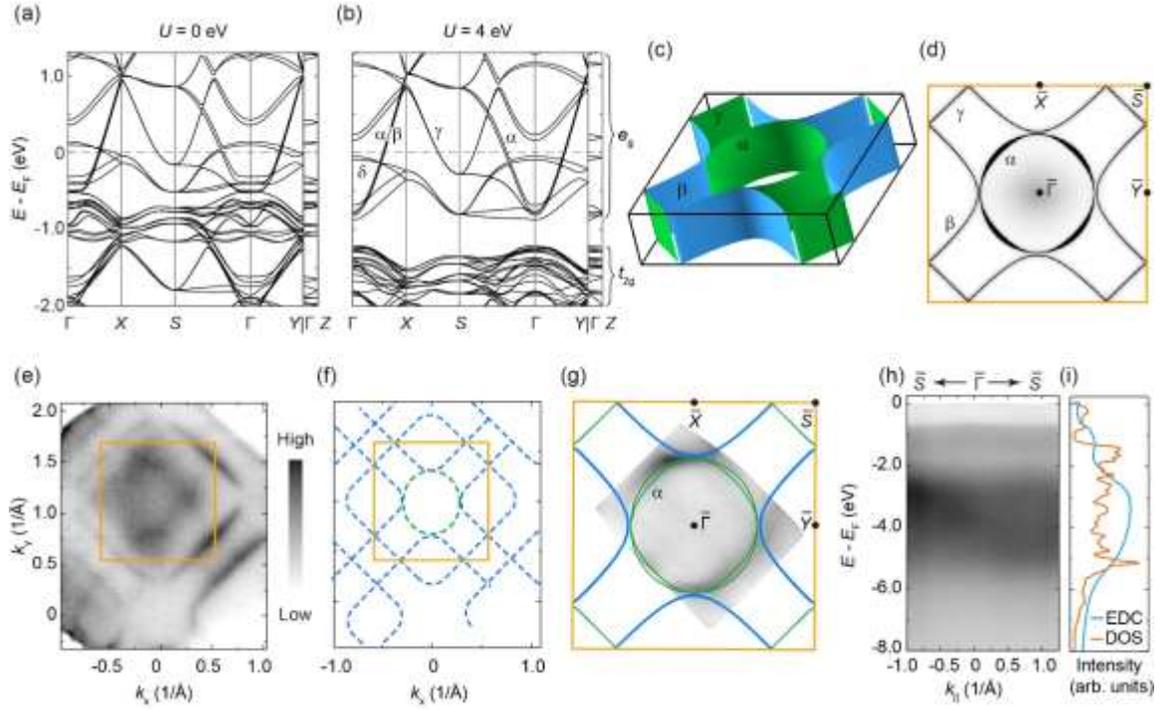

**Fig. 2.** Band structure of $La_3Ni_2O_7$. (a, b) Calculated band structure at the ambient pressure with onsite Coulomb interaction on Ni atoms of $U = 0$ (a) and $U = 4$ eV (b). (c) Calculated three-dimensional (3D) Fermi surface (FS) shown in the conventional BZ. (d) Surface-projected FS calculation (surface-projected calculation of the band dispersions is shown in Supplementary). (e) FS map measured by integrating ARPES intensity in an energy window of 40 meV around $E_F$. (f) FS extracted from panel (e) according to ARPES intensity. The orange lines are the BZ of the conventional unit cell. (g) FS measured using a 7 eV laser at 40 K. ARPES intensity was integrated in an energy window of 40 meV around $E_F$. The calculated FS in the conventional BZ is overlaid for comparison. (h) Band dispersion measured in a large energy range along the $\overline{\Gamma S}$ direction. (i) Comparison between the integrated energy distribution curve (EDC) and calculated density of states (DOS). Data in (e), (h), and (i) were collected using linear-horizontally (LH) polarized photons at 75 eV at 18 K.

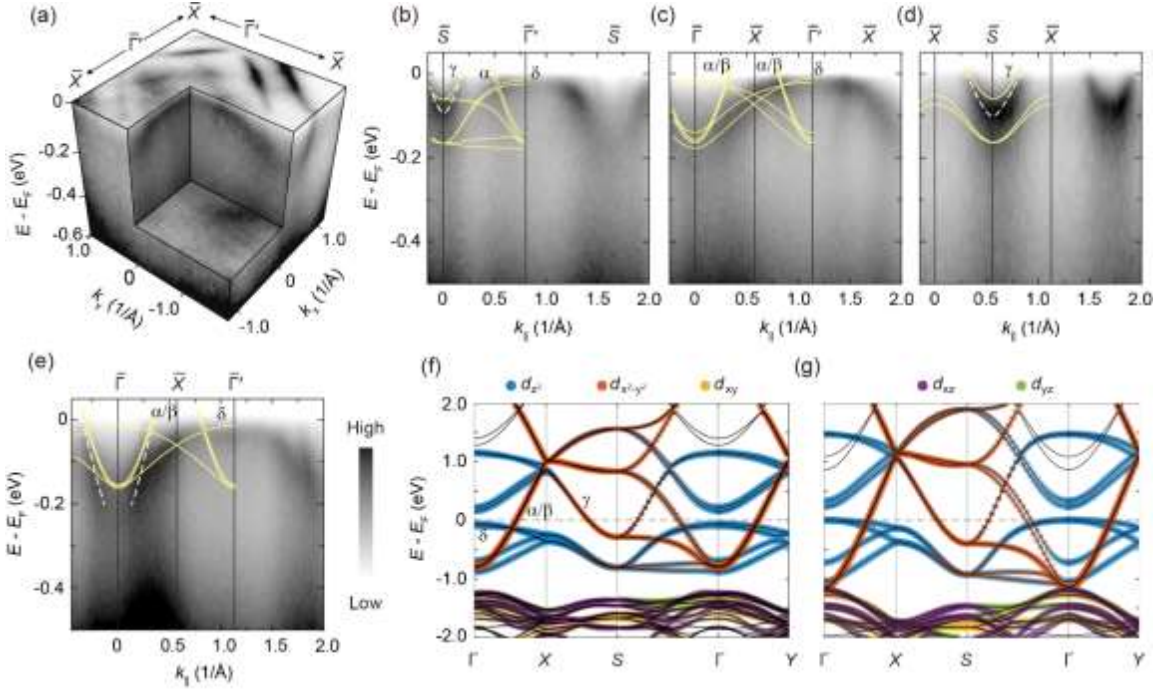

**Fig. 3.** Fine band dispersions along high-symmetry directions. (a) 3D plot of the electronic structure of La$_3$Ni$_2$O$_7$. (b-d), Band dispersions along the $\bar{\bar{\Gamma}}'\bar{S}$ (b), $\bar{\Gamma}\bar{X}$ (c), and $\bar{X}\bar{S}$ (d) directions. Data in (a)-(d) were collected using LH polarized photons at 75 eV at 18 K. (e) Band dispersions along $\bar{\Gamma}\bar{X}$ measured using LH polarized photons at 86 eV at 18 K. The yellow curves are calculated band structure for comparison, which are renormalized by a factor of 5. The white dashed lines are the guide to the eyes for the experimental band dispersions to indicate the orbital-selective renormalization factors. (f, g) Orbital-projected band structures at ambient pressure (f) and high pressure (g) with onsite Coulomb interaction $U = 4$ eV. The size of the circles represents the weight of the projected orbitals.

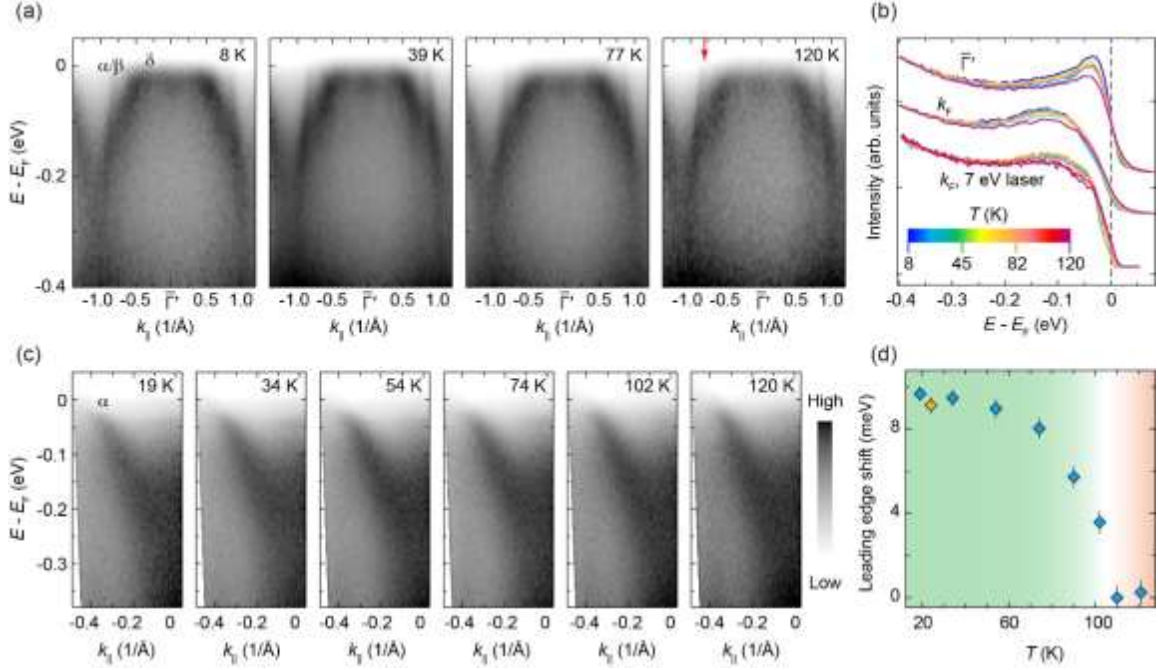

**Fig. 4.** Temperature-dependent ARPES measurements of $La_3Ni_2O_7$. (a) Band dispersions along $\bar{\Gamma}\bar{X}$ measured at selected temperatures. Data were collected using photons at 86 eV with LH polarization. (b) EDCs integrated around the $\bar{\Gamma}'$ point [data in (a)], $k_F = -0.8$ Å$^{-1}$ [the red arrow in (a)], and $k_F = 0.39$ Å$^{-1}$ [data in (c)] at selected temperatures. The EDCs are normalized by ARPES intensity in an energy window of 0.1 eV near -0.4 eV. (c) Band dispersions along $\bar{\Gamma}\bar{S}$ measured at selected temperatures. Data were collected using liner-vertically (LV) polarized laser at 7 eV. EDCs integrated near $k_F$ (red arrow) at selected temperatures are shown in panel (b). (d) EDC leading edge shift at $k_F = 0.39$ Å$^{-1}$ [data in (c)] with respect to the data at 110 K as a function of temperature. Aging effect has been excluded by temperature cycling measurement [the orange data point at 18 K in (d)].